\documentclass[prd,onecolumn,superscriptaddress]{revtex4}
\usepackage{amsmath}
\usepackage{adjustbox}
\usepackage{gensymb}
\usepackage{amsfonts}
\usepackage{CJKutf8}
\usepackage{graphicx}
\usepackage{hyperref}
\usepackage{amssymb}
\usepackage{cancel}
\usepackage{color}
\usepackage{amscd}
\usepackage{bm}
\usepackage{xcolor} 
\usepackage{adjustbox}
\makeatletter
\renewcommand{\maketag@@@}[1]{\hbox{\m@th\normalsize\normalfont#1}}%
\makeatletter
\begin{document}
\begin{CJK*}{UTF8}{gbsn}
\title{Extremalization approach to black hole thermodynamics: perturbations around higher-derivative gravities}

\author{Aonan Zhang}
\author{Qiang Wang}
\affiliation{Key Laboratory of High-precision Computation and Application of Quantum Field Theory of Hebei Province,
College of Physical Science and Technology, Hebei University, Baoding 071002, China}
\affiliation{Hebei Research Center of the Basic Discipline for Computational Physics, Baoding, 071002, China}
\author{Yong Xiao}
\email{xiaoyong@hbu.edu.cn}
\affiliation{Key Laboratory of High-precision Computation and Application of Quantum Field Theory of Hebei Province,
College of Physical Science and Technology, Hebei University, Baoding 071002, China}
\affiliation{Hebei Research Center of the Basic Discipline for Computational Physics, Baoding, 071002, China}
\affiliation{Higgs Centre for Theoretical Physics, School of Mathematics, University of Edinburgh, Edinburgh, EH9 3FD, United Kingdom}

\begin{abstract}
When higher-derivative terms are added to a gravitational action, black hole solutions and their thermodynamic properties are generally corrected. Recent progress has shown that, by treating higher-derivative operators as perturbations, the first-order corrections to black hole thermodynamics can be obtained without explicit knowledge of the corresponding perturbed black hole solutions. This result can be understood as a consequence of an extremalization principle underlying the Euclidean action formulation of black hole thermodynamics. In this paper, we emphasize that this extremalization approach is not restricted to perturbations around Einstein gravity. Instead, it can be applied to perturbations of more general higher-derivative gravity theories whose black hole solutions are already known and can be taken as the zeroth-order background. As an explicit illustration, we consider Einstein--Gauss--Bonnet gravity as the zeroth-order theory and study the first-order thermodynamic corrections induced by further  higher-order curvature operators. We show that these corrections can be derived without solving the perturbed black hole solutions, both in asymptotically flat and asymptotically AdS spacetimes.
\end{abstract}

 \maketitle
 
\section{Introduction}
Einstein's general relativity is described by the Einstein-Hilbert action. However, from the perspective of effective field theory, we can add higher-derivative terms to it, so the Lagrangian can be written as:
\begin{align}
    L=\frac{1}{16\pi}(R-2\Lambda)+\sum_m \gamma_m L_m, \label{lag1}
\end{align}
where $\gamma_m$ are the couplings of the higher-derivative terms $L_m$. In practice, such higher-derivative terms can be obtained by integrating out small-scale matter field fluctuations or derived from the low-energy effective action of string theory via a top-down approach \cite{Boulware:1985wk,Cardoso:2018ptl,Clifton:2011jh}. These higher-derivative terms may induce non-trivial effects on black hole physics; see, e.g., Refs.~\cite{Borah:2025crf, Borah:2025tvw}.

In general, considering the perturbations due to  the higher-derivative terms in \eqref{lag1}, the known black hole solutions and black hole thermodynamics in Einstein gravity will receive higher-order corrections. In fact, perturbed black hole solutions are often difficult to obtain, especially for rotating black holes. Surprisingly, at the first-order level in $\gamma_m$, the perturbed black hole thermodynamic quantities can be derived without knowing the perturbed black hole solutions. This phenomenon was first observed by Gubser et al. in 1999, but it was regarded as a puzzling phenomenon  at that time and lacked a proper theoretical explanation \cite{Gubser:1998nz,Caldarelli:1999ar,Landsteiner:1999gb}. 

Only recently, this fact was rediscovered by Reall and Santos in their study of Kerr black holes, who provided a reasonable explanation \cite{Reall:2019sah}. Therefore, subsequent literature often refers to this method as the Reall-Santos method. For specific higher-derivative terms, they obtained thermodynamic corrections accurate to $\mathcal{O}(\gamma_m)$, and these thermodynamic corrections are non-perturbative in $J/M^2$ \cite{Reall:2019sah}. In contrast, existing techniques for solving the rotating black hole metric induced by higher-derivative corrections can only handle the perturbative expansion in $J/M^2$ \cite{Cano:2019ore}, so the thermodynamic quantities derived based on the black hole metric can only yield perturbative results in $J/M^2$. Thus, this implies that, in such scenarios, the Reall-Santos method has an irreplaceable advantage, obtaining results that are unattainable by conventional methods.

This method is of great significance because these thermodynamic corrections themselves contain important physical information, which has stimulated many excellent works in recent years \cite{Ma:2023qqj,Hu:2023gru,Ma:2024ynp,Xiao:2022auy,Xiao:2023two,Xiao:2025llj,Guo:2025ohn,Guo:2025muo,Chen:2025ary}. On one hand, it can be used to precisely verify AdS/CFT \cite{Gubser:1998nz,Ma:2024ynp}. In fact, the motivation of Gubser et al. in \cite{Caldarelli:1999ar} was to understand the famous 3/4 discrepancy between the free energy of weakly coupled CFT and that of strongly coupled CFT calculated via holographic duality by introducing higher-derivative terms. On the other hand, the corrected physical quantities can reveal important quantum gravity effects. For example, the increase or decrease in the mass of extremal black holes at fixed $Q$ after switching on the couplings $\gamma_m$ can be used to verify the weak gravity conjecture (WGC); or conversely, WGC can provide physical constraints on the possible higher-derivative terms \cite{Xiao:2025llj,Cremonini:2019wdk,Harlow:2022ich}.

In this paper, we focus on the method itself. The previous works have all studied the corresponding correction behaviors based on Einstein gravity. However, the extremalization principle underlying this method is highly universal, meaning the method has great extensibility. In particular, we emphasize that based on the same spirit, for other gravitational theories, as long as their black hole solutions are known, we can take them as the zeroth-order results and further investigate the first-order perturbative thermodynamics induced by even higher-order curvature operators.

The structure of the remainder of this paper is as follows. In the second section, we briefly review the Reall-Santos method and then illustrate how to extend this method to more general higher-order gravities using its underlying extremalization idea. In the third section, we take the black hole thermodynamics of Einstein-Gauss-Bonnet gravity as an example, treating it as the zeroth-order starting point, and discuss the corrections to the thermodynamic quantities in asymptotically flat and asymptotically AdS spacetimes in each subsection. The subtleties related to asymptotically AdS spacetimes are carefully handled. The last section is the conclusion.

\section{A Review of the Extremalization  Method and Its Generalization}

In this section, we present a concise review of the method, clarify that its essence lies in perturbation analysis near extremal points, and discuss its potential generalizations.

\subsection{A Review of the Extremalization Method}

To illustrate the idea of the method, we start with the asymptotically flat, spherically symmetric case ($\Lambda=0, J=0$). For the system described by the Lagrangian $L=\frac{1}{16\pi} R+\gamma L_{\text{hd}}$, we first consider its black hole solutions. We denote the solution of the original Einstein-Hilbert action, i.e., the original Schwarzschild solution, as $\bar{g}_{\mu\nu}$. Correspondingly, the perturbed black hole solution induced by the higher-derivative terms is denoted as $g_{\mu\nu}$. At the first-order perturbation level of interest in this paper, we have the following relation:
\begin{align}
    g_{\mu\nu}(T) = \bar{g}_{\mu\nu}(T) + \mathcal{O}(\gamma).
\end{align}
We require the unperturbed metric and the perturbed black hole metric to have the same temperature $T$, and the reason will become clear shortly.

To analyze black hole thermodynamics, we should compute the Euclidean action \cite{hawkingpage, Dutta:2006vs, Gibbons:2004ai}, which consists of two parts: contributions from the Einstein-Hilbert term and the higher-derivative terms:
\begin{align}
    I_{\text{tot}} = I_{\text{EH}} + \gamma I_{\text{hd}}.
\end{align}
We need to substitute the metric $g_{\mu\nu}$ into the calculation. For the first term, there is 
\begin{align}
    I_{\text{EH}}(g_{\mu\nu}(T)) = I_{\text{EH}}(\bar{g}_{\mu\nu}(T)) + \mathcal{O}(\gamma^2).
\end{align}
This is because $\bar{g}_{\mu\nu}$ is a solution to the original action and thus extremizes the functional $I_{\text{EH}}$. As we know, the inaccuracy caused by perturbations near the extremum only arises from higher-order terms of $\gamma$.

For the second term, we similarly have:
\begin{align}
    \gamma I_{\text{hd}}(g_{\mu\nu}(T)) = \gamma I_{\text{hd}}(\bar{g}_{\mu\nu}(T)) + \mathcal{O}(\gamma^2).
\end{align}
This is because there is already a coefficient $\gamma$ in front of this term, so the inaccuracy caused by the metric also appears at $\mathcal{O}(\gamma^2)$.

In summary, at the first-order accuracy level, there is no difference between using $g_{\mu\nu}(T)$ or $\bar{g}_{\mu\nu}(T)$ for calculations. Thus:
\begin{align}
    I_{\text{tot}}(g_{\mu\nu}(T)) \doteq I_{\text{tot}}(\bar{g}_{\mu\nu}(T)),
\end{align}
where we use $\doteq$ to indicate that this equality holds at the first-order level in $\gamma$. Therefore, we only need to know the unperturbed black hole metric $\bar{g}_{\mu\nu}(T)$ to obtain $I_{\text{tot}}(g_{\mu\nu}(T))$. Moreover, since $I_{\text{EH}}(\bar{g}_{\mu\nu}(T))$ is the Euclidean action of the original Schwarzschild black hole, we do not even need to perform any calculations for this part and can directly employ existing results from the literature. Once we get the complete expression of $I_{\text{tot}}$ and relate it to the partition function of the system, all the thermodynamic quantities can be deduced.

Here, we emphasize a key point when using this method. Mathematically, to make a reasonable comparison between the functionals $\mathcal{F}\{f_1(x)\}$ and $\mathcal{F}\{f_2(x)\}$, the two functions $f_1(x)$ and $f_2(x)$ need to share the same boundary conditions. Therefore, when applying the above method, the key is to ensure that the perturbed solution and the original unperturbed solution have the same asymptotic structure at infinity, including temperature $T$ (which is related to the imaginary time period), angular velocity $\Omega$, etc., depending on the thermal ensemble under study. In particular, for AdS spacetimes, we require both spacetimes to have the same effective cosmological constant $\Lambda_e$ (or cosmological radius $l_e$). Neglecting this point can easily lead to incorrect results \cite{Hu:2023gru, Xiao:2023two}.

\subsection{Generalization of the Method}
A careful examination of this method reveals that the idea of extremalization analysis is highly universal. We can even apply it to different physical disciplines beyond gravity. In Ref. \cite{Xiao:2023two}, we pointed out that this extremalization analysis actually provides a new perspective  for an assertion in perturbative quantum mechanics: the first-order correction to the energy can be obtained from the unperturbed quantum state. Furthermore, this extremalization analysis tells us that we are not limited to first-order corrections; in fact, $n$th-order corrections to physical quantities can be derived from $(n-1)$th-order solutions.

In this paper, we focus on another interesting aspect: previous applications of this method have all involved perturbation calculations around the Einstein-Hilbert action. In fact, considering the universality of the extremalization idea, we can also investigate first-order perturbations around other gravitational models. For example, we can consider the following Lagrangian:
\begin{align}
    L = L_0 + \gamma L_{\text{ehd}},
\end{align}
where
\begin{align}
    L_0 = \frac{R}{16\pi} + \alpha L_{\text{hd}},
\end{align}
includes both the two-derivative Einstein-Hilbert term $\frac{R}{16\pi}$ and the four-derivative term $\alpha L_{\text{hd}}$, and $L_{\text{ehd}}$ denotes even higher-derivative terms. If we have obtained the corresponding black hole solutions of the system described by $L_0$, we can take these results as the zeroth-order approximation and treat $\gamma L_{\text{ehd}}$ as a perturbation. 

We still denote the original black hole solution and the first-order perturbed black hole solution as $\bar{g}_{\mu\nu}$ and $g_{\mu\nu}$, respectively, with the relation  $g_{\mu\nu}=\bar{g}_{\mu\nu}+\mathcal{O}(\gamma)$. To analyze black hole thermodynamics, we write the Euclidean action as:
\begin{align}
    I_{\text{tot}}= I_0 + \gamma I_{\text{ehd}}.
\end{align}
Repeating the previous analysis, we know that at the first-order accuracy level in $\gamma$, we have:
\begin{align}
    I_{\text{tot}}(g_{\mu\nu}(T)) \doteq I_{\text{tot}}(\bar{g}_{\mu\nu}(T)).
\end{align}
That is, the first-order corrections to black hole thermodynamics induced by the $\gamma L_{\text{ehd}}$ term can be obtained without knowing the perturbed black hole metric.

A key feature of the corresponding thermodynamics is that it is accurate to first-order in $\gamma$ but exact and non-perturbative in $\alpha$. Obviously, this allows us to gain more useful information about operator corrections and interactions between these operators.

\section{An Example: Quasi-Topological Gravity in $D=5$ Dimensions}

In the previous subsection, we have provided a general analysis of extending this extremalization method. Next, we illustrate it with a representative example. We will focus on cubic quasi-topological gravity, which is described by the Lagrangian:
\begin{align}
    L = L_0 + \gamma L_{\text{ehd}},
\end{align}
where
\begin{align}
    L_0 = \frac{1}{16\pi}\left[(R - 2\Lambda) + \alpha\left(R^2 - 4 R_{\mu\nu} R^{\mu\nu} + R_{\mu\nu\rho\sigma} R^{\mu\nu\rho\sigma}\right)\right], \label{egbl0}
\end{align}
\begin{align}
\mathcal{L}_{\text{ehd}}
&=
-\frac{7}{6}\,R_{ab}{}^{cd}\,R_{ce}{}^{bf}\,R_{df}{}^{ae}
- R_{cd}{}^{ab}\,R_{be}{}^{cd}\,R_{a}{}^{e}
-\frac{1}{2}\,R_{cd}{}^{ab}\,R_{a}{}^{c}\,R_{b}{}^{d}
\nonumber\\
&\quad
+\frac{1}{3}\,R_{a}{}^{b}\,R_{b}{}^{c}\,R_{c}{}^{a}
-\frac{1}{2}\,R\,R_{a}{}^{b}\,R_{b}{}^{a}
+\frac{1}{12}\,R^{3}\, . \label{ehd}
\end{align}
Specifically, the 4-derivative Lagrangian term is chosen as the well-known Gauss-Bonnet term. For the Gauss--Bonnet coupling $\alpha$, we follow the original solution in Ref.~\cite{Cai:2001dz}, where static spherically symmetric black holes in Einstein--Gauss--Bonnet gravity were derived. In that work, $\alpha$ is taken to be positive ($\alpha > 0$) to ensure a physically well-behaved vacuum and avoid ghost instabilities. On the other hand, the 6-derivative term is given by the combination \eqref{ehd}. According to the classification in Refs. \cite{Oliva:2010eb, Oliva:2010zd, Myers:2010ru}, in $D=5 $ dimensions, this is a unique combination that ensures the corresponding gravitational field equations contain at most second-order derivatives of the metric in the static, spherically symmetric case.

Next, we take $L_0$ as our zeroth-order part, which yields the conventional Einstein-Gauss-Bonnet black hole thermodynamics. We then treat $L_{\text{ehd}}$ as a perturbation and use the method described earlier to calculate the first-order thermodynamic corrections accurate to $\gamma$. We require the higher-derivative coupling $\gamma$ to be sufficiently small for the perturbation analysis to be valid.

\subsection{Zeroth-order Results}

As we mentioned, to use this convenient method for deriving thermodynamic corrections without solving the metric, we first need to be familiar with the zeroth-order results. Therefore, we briefly review the thermodynamics of Einstein-Gauss-Bonnet black holes. This gravity theory is described by $L_0$ \eqref{egbl0}, and its spherically symmetric black hole solution is \cite{Cai:2001dz,Cai:2003kt}:
\begin{align}
ds^2 = -\bar{f}(r) dt^2 + \frac{1}{\bar{g}(r)} dr^2 + r^2 d\Omega_3^2,
\end{align}
\begin{align}
\bar{f}(r) = \bar{g}(r) = 1 + \frac{r^2 \left(1 - \sqrt{1 + \frac{64 m \alpha}{3 \pi r^4} + \frac{4 \Lambda \alpha}{3}}\right)}{4 \alpha} \label{0thmetric}.
\end{align}
From this metric, the Hawking temperature of the black hole is obtained as:
\begin{align}
    \bar{T} = \frac{\sqrt{\bar{f}'(\bar{r}_h)\bar{g}'(\bar{r}_h)}}{4\pi} = \frac{3 \bar{r}_h - \Lambda \bar{r}_h^3}{24 \pi \alpha + 6 \pi \bar{r}_h^2} \label{0thtem}.
\end{align}
Note that we always use barred quantities to denote zeroth-order physical quantities  associated with $L_0$, while unbarred quantities are used for those including perturbations. Thus, $\bar{r}_h$ represents the outer horizon radius of the original unperturbed Einstein-Gauss-Bonnet black hole. In Eq. \eqref{0thmetric}, \( m \) denotes the mass parameter (not necessarily the physical mass $\bar{M}$) of the metric, which is determined by the condition \( f(\bar{r}_h) = g(\bar{r}_h) = 0 \); thus, it is related to the temperature through Eq. \eqref{0thtem}.

After Wick rotation, the Euclidean action can be calculated by:
\begin{align}
    I_E(\bar{g}_{\mu\nu}) = - \int_{\mathcal{M}}^{(Reg.)} L_0\, . \label{egbibar}
\end{align}
For AdS black holes, regularization is necessary to remove divergences and obtain a physically meaningful $I_E$.  We use the background subtraction method, as it is a convenient regularization technique, which subtracts the contribution of the background AdS spacetime from the black hole calculation results. Specifically, we define:
\begin{align}
    \int_{\mathcal{M}}^{(\text{Reg.})} \equiv \beta \lim_{r_c\rightarrow \infty} \bigl(\int_{\mathcal{V}_{r_c}}^{(\text{BH})} - \sqrt{\frac{g_{tt}^{\text{BH}}(r_c)}{g_{tt}^{\text{AdS}}(r_c)}} \int_{\mathcal{V}_{r_c}}^{(\text{AdS})}\bigr),
\end{align}
where $\beta \equiv 1/T$ is the period of Euclidean time, and $r_c$ is the infrared cutoff of the system. The prefactor of the second term is to ensure that the black hole spacetime and the reference spacetime have the same asymptotic geometry at infinity \cite{hawkingpage,Dutta:2006vs,Gibbons:2004ai,Xiao:2025icr,Guo:2025ohn}. An advantage of the background subtraction method is that, for pure gravity in AdS spacetime, boundary terms such as the Gibbons-Hawking-York term cancel out directly during background subtraction procedure, so we do not need to handle the cumbersome boundary terms corresponding to higher-derivative terms. Substituting the metric \eqref{0thmetric} into Eq. \eqref{egbibar} explicitly yields:
\begin{align}
 I_E(\bar{g}_{\mu\nu})= \bar{\beta} \bigg[ \frac{\pi  \left(24 \alpha ^2+\bar{r}_h^4-6 \alpha \bar{r}_h^2\right)}{8 \left(4 \alpha +\bar{r}_h^2\right)}+\frac{\pi  \Lambda  \left(\bar{r}_h^6+36 \alpha \bar{r}_h^4\right)}{48 \left(4 \alpha +\bar{r}_h^2\right)} 
\bigg]\label{0thie}.
\end{align}
The free energy is given by 
\begin{align}
    F = I_E/\beta, \label{0thfree}
\end{align}
and the entropy and mass can be derived from the formulas:
\begin{align}
    S = -\frac{\partial F}{\partial T}\quad \text{and} \quad M = F + T S.
\end{align}

Thus, for Einstein-Gauss-Bonnet gravity, we obtain:
\begin{align}
  & \bar{S} = \frac{1}{2} \pi^2 \bar{r}_h^3 + 6 \pi^2 \alpha \bar{r}_h, \label{0thentr} \\
  & \bar{M} = \frac{3 \pi  \bar{r}_h^2}{8} + \frac{3 \pi  \alpha }{4} - \frac{1}{16} \pi  \Lambda  \bar{r}_h^4. \label{0thmass}
\end{align}

In summary, the zeroth-order metric is given by Eq. \eqref{0thmetric}, and the zeroth-order thermodynamic quantities (temperature, Euclidean action, entropy, and mass) are respectively given by Eqs. \eqref{0thtem}, \eqref{0thie}, \eqref{0thentr}, and \eqref{0thmass}.

\subsection{First-Order Perturbation of Black Hole Thermodynamics: Asymptotically Flat Case}

We now evaluate the first-order corrections to Einstein-Gauss-Bonnet thermodynamics induced by $\gamma L_{\text{ehd}}$. As a warm-up, we start with the simpler asymptotically flat case. In this scenario, the Lagrangian is given by:
\begin{align}
L = L_0|_{\Lambda=0} + \gamma L_{\text{ehd}}.
\end{align}
At the zeroth-order level, the black hole metric reads:
\begin{align}
ds^2 = -\bar{f}(r) dt^2 + \frac{1}{\bar{g}(r)} dr^2 + r^2 d\Omega_3^2,
\end{align}
where
\begin{equation}
\bar{f}(r) = \bar{g}(r) = 1 + \frac{r^2 \left(1 - \sqrt{1 + \frac{64 m \alpha}{3 \pi r^4} }\right)}{4 \alpha} \label{sec2fg}.
\end{equation}
Treating $\gamma L_{\text{ehd}}$ as a perturbation of this system, the perturbed metric can be expressed as $g_{\mu\nu}(T) = \bar{g}_{\mu\nu}(T) + \mathcal{O}(\gamma)$.

The core insight of this paper is that, evaluating the Euclidean action $I_E(g_{\mu\nu})$ only requires knowledge of $\bar{g}_{\mu\nu}$. The action $I_E(g_{\mu\nu})$ consists of two components: $I_0(g_{\mu\nu})$ and $\gamma I_{\text{ehd}}$. Since $\bar{g}_{\mu\nu}$ extremizes the functional $I_0$, we have $I_0(g_{\mu\nu}) \doteq I_0(\bar{g}_{\mu\nu})$. Thus, for this part, we can directly adopt the known results for the unperturbed Einstein-Gauss-Bonnet black hole. The thermodynamic quantities for the asymptotically flat Einstein-Gauss-Bonnet black hole are obtained by taking the limit $\Lambda \rightarrow 0$ in Eqs. \eqref{0thie}, \eqref{0thtem}, \eqref{0thentr}, and \eqref{0thmass}. In particular, the Euclidean action is:
\begin{align}
I_0 \doteq \bar{\beta} \bigg[ \frac{\pi \left(24 \alpha^2 + \bar{r}_h^4 - 6 \alpha \bar{r}_h^2\right)}{8 \left(4 \alpha + \bar{r}_h^2\right)} \bigg] \label{ie0},
\end{align}
where $\bar{\beta} = 1/\bar{T}$. As emphasized in the previous section, to correctly apply this method, we require the perturbed and unperturbed black holes to share the same temperature $T$. Thus, we do not need to distinguish between $T$ and $\bar{T}$ explicitly, and the temperature satisfies the relation:
\begin{align}
T = \frac{\bar{r}_h}{8 \pi \alpha + 2 \pi \bar{r}_h^2} \label{sec2tem0},
\end{align}
Note that $\bar{r}_h$ here denotes the horizon radius of the unperturbed black hole. In principle, we could solve for $\bar{r}_h(T)$ from this equation to express the Euclidean action \eqref{ie0} as a function of $T$, but this is unnecessary. Instead, we can use $\bar{r}_h$ as an intermediate variable and apply the chain rule when performing calculations (e.g., computing $\frac{\partial F}{\partial T}$).

On the other hand, at the $\mathcal{O}(\gamma)$ accuracy level, $\gamma I_{\text{ehd}}(g_{\mu\nu}) \doteq \gamma I_{\text{ehd}}(\bar{g}_{\mu\nu})$, so we only need to substitute the zeroth-order metric \eqref{sec2fg} for the explicit calculation. This yields:
\begin{align}
\gamma I_{\text{ehd}} \doteq - \frac{\pi \gamma \left(4 \alpha + 7 \bar{r}_h^2\right)}{8 \bar{r}_h^2 \left(4 \alpha + \bar{r}_h^2\right)}.
\end{align}

Combining the two components, we obtain the Euclidean action $I_E = I_0 + \gamma I_{\text{ehd}}$. The free energy is then derived via $F = I_E/\beta$:
\begin{align}
F \doteq \frac{\pi \left(24 \alpha^2 + \bar{r}_h^4 - 6 \alpha \bar{r}_h^2\right)}{8 \left(4 \alpha + \bar{r}_h^2\right)} - \frac{\pi \gamma \left(4 \alpha + 7 \bar{r}_h^2\right)}{8 \bar{r}_h^2 \left(4 \alpha + \bar{r}_h^2\right)}.
\end{align}
The entropy and mass are obtained using $S = -\partial F/\partial T$ and $M = F + T S$, respectively:
\begin{align}
S &\doteq \frac{\pi^2 \bar{r}_h^3}{2} + 6 \pi^2 \alpha \bar{r}_h + \frac{\pi^2 \gamma \left(16 \alpha^2 + 7 \bar{r}_h^4 + 8 \alpha \bar{r}_h^2\right)}{2 \bar{r}_h^3 \left(\bar{r}_h^2 - 4 \alpha\right)}, \\
M &\doteq \frac{3 \pi \bar{r}_h^2}{8} + \frac{3 \pi \alpha}{4} + \frac{\pi \gamma \left(12 \alpha + 7 \bar{r}_h^2\right)}{8 \bar{r}_h^2 \left(\bar{r}_h^2 - 4 \alpha\right)}.
\end{align}
Through the relation $\bar{r}_h(T)$, all the quantities are effectively expressed as functions of temperature $T$.

It is worthy to note that there is a technique to rewrite these formulas in terms of the perturbed horizon radius $r_h$. Starting from the Lagrangian, we can compute the black hole entropy using the Wald entropy formula \cite{Wald:1993nt,Iyer:1994ys}:
\begin{equation}
S_{\text{W}} = 2\pi \oint_{\mathcal{H}} \frac{\partial \mathcal{L}}{\partial R_{abcd}} \, \epsilon_{ab} \epsilon_{cd}, \label{swformula}
\end{equation}
where $\epsilon_{ab}$ is the binormal vector of the horizon $\mathcal{H}$. It is well known that the contribution of the Einstein-Hilbert sector to the Wald entropy is always $A/4$, where $A = 2\pi^2 r_h^3$ is the horizon area in $D=5$. Actually, for any spherically symmetric solution, the Einstein-Gauss-Bonnet contribution to the Wald entropy can be written as $\frac{1}{2} \pi^2 r_h^3 + 6 \pi^2 \alpha r_h$, independent of the specific metric form. At $\mathcal{O}(\gamma)$, the contribution from $\gamma L_{\text{ehd}}$ can be directly computed using the zeroth-order metric, again due to the presence of the prefactor $\gamma$ \footnote{In fact, given the quasi-topological nature of the Lagrangian in this specific example, the Wald entropy can be explicitly obtained as a function of $r_h$ independent of the concrete form of the spherically symmetric black hole solution. In other words, Eq. \eqref{swald} is an exact result for any order in $\gamma$.}. Thus, from the formula \eqref{swformula}, we readily obtain:
\begin{align}
S_W \doteq (\frac{1}{2} \pi^2 r_h^3 + 6 \pi^2 \alpha r_h ) + \frac{3 \pi^2 \gamma}{2 r_h} \label{swald}.
\end{align}
Equating this to the previously derived entropy $S(\bar{r}_h)$, we can solve for the relation $\bar{r}_h \doteq r_h(1 + \gamma b(r_h))$:
\begin{align}
\bar{r}_h \doteq r_h - \frac{4 \gamma \left(r_h^2 + \alpha\right)}{3 r_h^3 \left(r_h^2 - 4 \alpha\right)}.
\end{align}
Substituting this back, all physical quantities can be expressed in terms of $r_h$. The resulting free energy, mass, and temperature are:
\begin{align}
T &\doteq \frac{r_h}{8 \pi \alpha + 2 \pi r_h^2} + \frac{2 \gamma \left(\alpha + r_h^2\right)}{3 \pi r_h^3 \left(4 \alpha + r_h^2\right)^2} ,\\
F &\doteq \frac{\pi \left(24 \alpha^2 + r_h^4 - 6 \alpha r_h^2\right)}{8 \left(4 \alpha + r_h^2\right)} - \frac{\pi \gamma \left(4 \alpha + 7 r_h^2\right)}{8 r_h^2 \left(4 \alpha + r_h^2\right)}, \\
M &\doteq \frac{3 \pi \alpha}{4} + \frac{3 \pi r_h^2}{8} - \frac{\pi \gamma}{8 r_h^2}\, .
\end{align}
In summary, we have obtained the thermodynamic quantities accurate to $\mathcal{O}(\gamma)$ without explicitly solving for the perturbed black hole solution.

\subsection{First-Order Perturbation of Black Hole Thermodynamics: Asymptotically AdS Case}

We next consider the asymptotically AdS case. The perturbed black hole metric induced by the perturbation $\gamma L_{\text{ehd}}$ takes the form:
\begin{align}
ds^2 = -f(r) dt^2 + \frac{1}{g(r)} dr^2 + r^2 d\Omega_3, \label{metricxx}
\end{align}
At infinity, its behavior tends to a pure AdS spacetime, i.e.,
\begin{align}
f(r) \sim 1 - \frac{\Lambda_e r^2}{6}. \label{asymf}
\end{align}
Note that $\Lambda_e$ is the effective cosmological constant appearing in the metric, while $\Lambda$ is the bare cosmological constant term of the theory in the Lagrangian. The subtlety here is that, due to the presence of higher-derivative terms, $\Lambda_e$ is generally not equal to $\Lambda$. As emphasized earlier, for a reasonable comparison between the perturbed and unperturbed metric, we need to require them to have the same asymptotic geometry at infinity. Therefore, we should first clarify the asymptotic behavior of the spacetime (though, as always, we don't need to know the complete explicit form of Eq.\eqref{metricxx}.).

In our previous work \cite{Xiao:2023two}, we derived a formula to calculate $\Lambda_e$:
\begin{equation}
\Lambda_e - \Lambda = \sum_{k} \frac{k - D}{2D} \alpha_k \mathcal{L}_k^{(\text{AdS}_{\Lambda_e})},
\end{equation}
where $D$ denotes the spacetime dimension, $k$ counts the number of derivatives acting on the metric in the Lagrangian $\mathcal{L}_k$, $\mathcal{L}_k^{(\text{AdS})}$ is its value on pure AdS, and $\alpha_k$ are the corresponding coupling parameters.

For the metric of the original Einstein-Gauss-Bonnet black hole (Eq. \eqref{0thmetric}), its asymptotic behavior at infinity is:
\begin{align}
\bar{f}(r) \sim \frac{r^2 \left(1 - \sqrt{\frac{4 \alpha \Lambda}{3} + 1}\right)}{4 \alpha},
\end{align}
while after adding the higher-derivative term $\gamma L_{\text{ehd}}$, at $\mathcal{O}(\gamma)$, we have:
\begin{align}
f(r) \sim \frac{r^2 \left(1 - \sqrt{\frac{4 \alpha \Lambda}{3} + 1}\right)}{4 \alpha} \bigg(1 + \frac{\gamma \left(1 + \frac{2 \alpha \Lambda}{3} - \sqrt{1 + \frac{4 \alpha \Lambda}{3}}\right)}{24 \alpha^2 \sqrt{1 + \frac{4 \alpha \Lambda}{3}}} \bigg).
\end{align}
Obviously, $f(r)$ and $\bar{f}(r)$ have different asymptotic behaviors at infinity. Thus the important point here is that we cannot directly take Eq. \eqref{0thmetric} as our zeroth-order metric, as this violates the requirement of identical asymptotic geometry. Hence, we need to rescale the time coordinate. We choose the following metric as the unperturbed zeroth-order metric:
\begin{align}
ds^2 = -\lambda_0^2 \bar{f}(r) dt^2 + \frac{1}{\bar{g}(r)} dr^2 + r^2 d\Omega_3 \label{red0thmetric},
\end{align}
where the constant rescaling factor $\lambda_0$ is given by:
\begin{align}
\lambda_0^2 = 1 + \frac{\gamma \left(1 + \frac{2 \alpha \Lambda}{3} - \sqrt{1 + \frac{4 \alpha \Lambda}{3}}\right)}{24 \alpha^2 \sqrt{1 + \frac{4 \alpha \Lambda}{3}}}.
\end{align}
Since only the time coordinate is rescaled, this metric remains a solution to the Einstein-Gauss-Bonnet action and satisfies our requirement of the same asymptotic geometry at infinity. (Besides, when $\Lambda_e \neq \Lambda$, apart from adopting the redshifted zeroth-order metric to handle the problem, an equivalent action decomposition method can also be used, see the appendix of \cite{Xiao:2023two} and related works \cite{Guo:2025muo, Chen:2025ary}.)

For the black hole metric \eqref{red0thmetric}, its time coordinate is rescaled compared to the familiar metric \eqref{0thmetric}, so its temperature is correspondingly rescaled to be  $\lambda_0 \bar{T}$, where $\bar{T}$ is given in Eq.\eqref{0thtem}. Now we require the perturbed and unperturbed black holes to share the same temperature, leading to:
\begin{align}
T = \lambda_0 \bigg(\frac{3 \bar{r}_h - \Lambda \bar{r}_h^3}{24 \pi \alpha + 6 \pi \bar{r}_h^2} \bigg).
\end{align}
Once the zeroth-order metric is chosen, the subsequent calculations are the same as in the previous section. First, based on the extremalization principle, the $I_0$ part can be borrowed from the Euclidean action of  the metric \eqref{red0thmetric}, and the corresponding free energy (differing from that of metric \eqref{0thmetric} only by a redshift factor $\lambda_0$) is:
\begin{align}
F_0 \doteq \lambda_0 \bigg[ \frac{\pi \left(24 \alpha^2 + \bar{r}_h^4 - 6 \alpha \bar{r}_h^2\right)}{8 \left(4 \alpha + \bar{r}_h^2\right)} + \frac{\pi \Lambda \left(\bar{r}_h^6 + 36 \alpha \bar{r}_h^4\right)}{48 \left(4 \alpha + \bar{r}_h^2\right)} \bigg].
\end{align}
For the second part $\gamma I_{\text{ehd}}$, we need to substitute the zeroth-order metric \eqref{red0thmetric} into the following integral:
\begin{align}
\gamma I_{\text{ehd}} = \gamma \int_{\mathcal{M}}^{(Reg.)} L_{\text{ehd}}\, .
\end{align}
The corresponding free energy is calculated as
\begin{align}
\begin{split}
\gamma F_{\text{ehd}} \doteq & \frac{\lambda_0 \, \pi \gamma}{2304 \alpha^2 \bar{r}_h^2 \left(4 \alpha + \bar{r}_h^2\right)} \bigg( -3 \left(384 \alpha^3 + \Lambda \bar{r}_h^8 + \bar{r}_h^6 (4 \alpha \Lambda - 6) - 12 \alpha \bar{r}_h^4 (16 \alpha \Lambda + 3) + 624 \alpha^2 \bar{r}_h^2\right) \\
& + \frac{\bar{r}_h^2 (2 \alpha \Lambda + 3) \left(4 \alpha + \bar{r}_h^2\right) \left(\Lambda \bar{r}_h^4 - 12 \alpha - 6 \bar{r}_h^2\right)}{\sqrt{\frac{4 \alpha \Lambda}{3} + 1}} \bigg).
\end{split}
\end{align}
Combining the two parts, we obtain the total free energy $F = F_0 + \gamma F_{\text{ehd}}$. The black hole entropy is then calculated from $S=-\frac{\partial F}{\partial T}$ and given by
\begin{align}
\begin{split}
S \doteq & \left(\frac{\pi^2 \bar{r}_h^3}{2} + 6 \pi^2 \alpha \bar{r}_h\right) + \frac{\pi^2 \gamma}{96 \alpha^2 \bar{r}_h^3 \sqrt{4 \alpha \Lambda + 3} \left(-12 \alpha + \Lambda \bar{r}_h^4 + 3 \bar{r}_h^2 (4 \alpha \Lambda + 1)\right)} \bigg( \sqrt{3} \bar{r}_h^4 (2 \alpha \Lambda + 3) \left(4 \alpha + \bar{r}_h^2\right)^2 \left(\Lambda \bar{r}_h^2 - 3\right) \\
& + 3 \sqrt{4 \alpha \Lambda + 3} \left(768 \alpha^4 - \Lambda \bar{r}_h^{10} + \bar{r}_h^8 (3 - 8 \alpha \Lambda) - 8 \alpha \bar{r}_h^6 (2 \alpha \Lambda - 3) + 384 \alpha^2 \bar{r}_h^4 (\alpha \Lambda + 1) + 384 \alpha^3 \bar{r}_h^2\right) \bigg).
\end{split}
\end{align}
The mass can be derived from $M = F + T S$. However, these expressions are relatively complex when using $\bar{r}_h$ as the variable. Utilizing the technique from the previous section, we note that the Wald entropy is still given by the form \eqref{swald}. Equating $S_W(r_h) \doteq S(\bar{r}_h)$, we solve for the relation $\bar{r}_h \doteq r_h (1+ \gamma b(r_h))$. Substituting it back into the expressions for temperature, mass and free energy, we find the final results become: 
\begin{align}
T &\doteq \frac{3 r_h - \Lambda r_h^3}{24 \pi \alpha + 6 \pi r_h^2} + \gamma \frac{4 \left(\alpha + r_h^2\right) - \Lambda r_h^4}{6 \pi r_h^3 \left(4 \alpha + r_h^2\right)^2};
\end{align}
\begin{align}
M &\doteq \bigg(\frac{3 \pi r_h^2}{8} + \frac{3 \pi \alpha}{4} - \frac{1}{16} \pi \Lambda r_h^4\bigg) - \frac{\pi \gamma}{8 r_h^2} ;
\end{align}
\begin{align}
F\doteq \bigg( \frac{\pi \left(24 \alpha^2 +  r_h^4 - 6 \alpha  r_h^2\right)}{8 \left(4 \alpha +  r_h^2\right)} + \frac{\pi \Lambda \left( r_h^6 + 36 \alpha  r_h^4\right)}{48 \left(4 \alpha +  r_h^2\right)} \bigg)   +\frac{\pi  \gamma \left(-144 \alpha ^2+8 \Lambda   r_h^6+ r_h^4 (48 \alpha  \Lambda -29)-200 \alpha   r_h^2\right)}{24 \left( r_h^3+4 \alpha   r_h\right)^2}.
\end{align}

To verify the correctness of our method, we compare our results with those in Ref. \cite{Myers:2010ru} which solved cubic quasi-topological gravity and studied black hole thermodynamic quantities directly from the metric. We find that our results are identical to theirs at $\mathcal{O}(\gamma)$ that we are interested in. When making the comparison, note the differences in conventions. Ref. \cite{Myers:2010ru} uses coupling parameters $\lambda$ and $\mu$, which are related to our parameters by $\lambda = \frac{2 \alpha}{L^2}$ and $\mu = \frac{\gamma}{3 L^4}$. In addition, our convention of the black hole behavior \eqref{asymf}  corresponds to $f_\infty = 1$ in their work.

\section{Concluding Remarks}

Previously, when using the extremalization method to analyze black hole thermodynamic problems, Einstein gravity was always treated as the zeroth-order approximation, with thermodynamic corrections induced by higher-derivative terms investigated based on this foundation. In this paper, we point out that we can take a solved gravitational theory (not necessarily Einstein gravity) as the zeroth-order approximation and consider perturbations on this basis, and show that this method remains applicable.

We illustrated this with quasi-topological gravity in five dimensions as an example. The asymptotically flat case serves as a relatively straightforward example, while the asymptotically AdS case involves some subtleties. In fact, there has been controversy when applying the method to asymptotically AdS spacetimes (see Refs. \cite{Xiao:2023two, Hu:2023gru}). In particular, our prior work \cite{Xiao:2023two} highlighted the importance of boundary conditions: only by aligning the asymptotic geometries of the perturbed and unperturbed metrics at infinity, can we correctly apply this method. This insight has since been successfully utilized in studies on higher-derivative corrections to Kerr-AdS thermodynamics in  \cite{Guo:2025muo, Chen:2025ary}. Accordingly, another important aspect of the present work is that it demonstrates how to correctly handle thermodynamic corrections of AdS black holes in scenarios more complex than Einstein gravity, thereby removing conceptual and technical obstacles for future related research.

Although this work focuses on static spherically symmetric black holes, the extremalization method is in principle extendable to rotating black holes. For instance, the method has been successfully demonstrated for Kerr \cite{Reall:2019sah} and Kerr-AdS black holes \cite{Guo:2025muo, Chen:2025ary}, with Einstein gravity as the zeroth-order theory. In our generalized framework, the zeroth-order theory can be an arbitrary higher-derivative gravity. The only issue is that rotating black hole solutions are seldom known in these theories. If such solutions are available and their corresponding thermodynamics can be clearly analyzed in future work, our method can be applied smoothly.

We mentioned that $n$th-order thermodynamic corrections can be derived from $(n-1)$th-order solutions. This implies that we can obtain thermodynamic corrections at $\mathcal{O}(\gamma^2)$ once the black hole metric at $\mathcal{O}(\gamma)$ is known, and so on. Anyway, we emphasize that the primary utility of the extremalization method is in cases where constructing the next-leading-order black hole solutions is challenging.

\section*{Acknowledgements}

YX is grateful to the Higgs Centre for Theoretical Physics at the University of Edinburgh for providing research facilities and hospitality during the visit. This work was supported in part by the National Natural Science Foundation of China (Grant No. 12475048), the Hebei Natural Science Foundation (Grant No. A2024201012), the Science Research Project of Hebei Education Department (Grant No. JCZX2026019), and the China Scholarship Council (Grant No. 202408130101).

\end{CJK*}

\end{document}